\documentclass[aps,twocolumn,showpacs,preprintnumbers,amsmath,amssymb,pre]{revtex4-1}
\usepackage{graphicx}
\usepackage{bm}
\usepackage{color,ulem}

\begin{document}

\title{Molecular dynamics on nonequilibrium motion of a colloidal 
particle driven by an external torque}

\author{Donghwan Yoo$^1$}
\author{Youngkyun Jung$^2$}
\author{Chulan~Kwon$^1$} \email{ckwon@mju.ac.kr}
\affiliation{$^1$Department of Physics, Myongji University, Yongin, Gyeonggi-Do,
17058,  Korea}
\affiliation{$^2$ National Institute of Supercomputing and Networking, Korea Institute of Science and Technology Information, Daejeon, 34141, Korea}

\date{today}

\begin{abstract}
We investigate the motion of a colloidal particle driven out of equilibrium by 
an external torque. We use molecular dynamics simulation as an alternative to the Langevin dynamics. We prepare a heat bath composed of thousands of particles interacting with each other through the Lennard-Jones potential and impose the Langevin thermostat to maintain heat bath in equilibrium. We 
consider a single colloidal particle to interact with the particles of the heat bath 
also by the Lennard-Jones potential, without applying any types of dissipative and fluctuating forces used in the Langevin dynamics. We set up simulation protocol fit for the overdamped limit as in real experiments by increasing the size and mass of the colloidal particle. We study nonequilibrium fluctuations for work and heat produced incessantly in time and compare results with those obtained from the previous studies via the overdamped Langevin dynamics. We confirm the Gallavotti-Cohen symmetry and the fluctuation theorem for the work production. 

\end{abstract}
\maketitle

\section{Introduction}

Nonequilibrium thermal fluctuations for a small system in contact with heat bath in equilibrium become very large and exhibit interesting properties that are universal over different systems. The underlying principle for the universal properties is the fluctuation theorem (FT), which was first discovered for a deterministic system thermostatted so as to conserve kinetic energy~\cite{evans93,evans94,gallavotti}.  The FT was later proven to hold for a wide class of stochastic systems~\cite{jarzynski,crooks,kurchan,lebowitz}. It deals with the thermal fluctuations of thermodynamic quantities produced persistently in time, such as work and heat in a nonequilibrium process. A typical form of the FT is given by $\langle e^{-\beta tw}\rangle=1$, where $w$ is the rate of work production piled over a period $t$ in nonequilibrium dynamics driven by a nonconservative force acting on the system initially prepared in equilibrium with heat bath. $\beta$ is the inverse temperature $(k_B T)^{-1}$ for the Boltzmann constant $k_B$ and the temperature $T$ of the heat bath. The bracket denotes the average over all possible fluctuations. To investigate the FT and related issues, it is required to accurately deal with the probability distribution of the thermal fluctuations in a nonequilibrium process.

Work and heat are path-dependent quantities. The thermal fluctuations of such quantities arise from many different paths (trajectories) along which the system evolves in time. One essentially needs the ensemble average over all paths from the past to the present, unlike the usual ensemble average over all states at a certain time. The path integral theory to deal with trajectory-dependent fluctuations was developed by Onsager and Mathlup based on the Langevin equation~\cite{onsager}, and was used successfully to prove the FT~\cite{kurchan}.    

In the experimental side, the fluctuations around the average value of a thermodynamic quantity for a small system become so predominant that they are measurable with high accuracy by using modern technologies.  Interesting experiments were carried out to confirm the FT for various systems such as a colloidal particle in a moving optical trap~\cite{wang,trepagnier}, a molecule in the AFM pulled by an external force~\cite{hummer}, an electrical circuit driven by a small current~\cite{garnier}, a harmonic oscillator under an external force~\cite{douarche,joubaud}, an RNA molecule unfolded and refolded by optical tweezers~\cite{liphardt,collin}, a rotating motor protein $\textrm{F}_1$-ATPase~\cite{hayashi}, and a colloidal particle in breathing harmonic potential~\cite{pak}. 

An alternative approach to the study on large fluctuations for a small system is molecular dynamics (MD). MD deals with both a small system and a heat bath consisting of an extremely high number of molecules by taking into account the interactions in detail among them, and it mimics an experiment as realistically as possible. It plays the role of the bridge between experiment and phenomenological theory such as the Langevin dynamics. The first MD study in this field was carried out accompanying an experiment for a colloidal particle in a moving optical trap~\cite{wang}. Recently, MD studies were done to confirm the FT for non-equilibrium motions such as the effusion of an ideal gas through a hole between two compartments~\cite{cleuren} and the motion of a particle in a moving billiard~\cite{schmick}. 

In our work, we investigate via MD simulation the motion of a colloidal particle in a liquid driven out of equilibrium by an external non-conservative force generating torque. Such external torque was found to produce non-zero circulating current even in steady state, characterizing nonequilibrium steady state~\cite{kwon-ao-thouless}. Energetically, it was found to produce work and heat incessantly in time~\cite{kwon,noh}. Such system was suggested as a heat engine~\cite{fillinger} and investigated as a model system to examine optimal efficiency for maximum power~\cite{jd_engine}. We examine and confirm various nonequilibrium properties including the FT. We compare the results from the MD simulation with those found from the Langevin dynamics~\cite{kwon,noh}.

\section{Nonequilibrium Molecular dynamics}

The particles employed in the MD are composed of molecules in liquid and the colloidal particle immersed in them. The liquid plays the role of a heat bath kept in equilibrium at a fixed temperature $T$. Liquid molecules are designed to interact with each other via the Lennard--Jones (LJ) potential:
\begin{equation}
V_{\textrm{LJ}}(r;\sigma)=4\epsilon\left[ \left(\frac{\sigma}{r}\right)^{12}-\left(\frac{\sigma}{r}\right)^{6}\right]~,
\label{LJ_potential}
\end{equation}
where $\sigma$ and $\epsilon$ define the effective diameter (radius of cross section) of a pair of molecules and the intensity of interactions, respectively. To maintain the liquid in equilibrium, we use the Langevin thermostat in which the same sort of dissipative and fluctuating forces are used as in the Langevin equation~\cite{grest}. Then, the equation of motion of the $i$-th molecule with mass $m$ for position $\mathbf{x}_i$ and velocity  $\mathbf{v}_i$ are given  as  
\begin{eqnarray}
m\dot{\mathbf{v}}_i&=&-{\sum_{j(\neq i)}}\boldsymbol{\nabla}_iV_{\textrm{LJ}}(|\mathbf{x}_i-\mathbf{x}_j|; \sigma)-\gamma\mathbf{v}_i +\sqrt{\gamma \beta^{-1}}\boldsymbol{\xi}_i(t)\nonumber\\
&&~~-\boldsymbol{\nabla}_iV_{\textrm{LJ}}(|\mathbf{x}_i-\mathbf{x}|; \sigma')~,
\label{reservoir}
\end{eqnarray}
where $\mathbf{x}$ is the position of the colloidal particle. For the LJ potential between the colloidal particle and a liquid molecule, a different parameter $\sigma'$ is used with the same $\epsilon$. $\sigma'$ is the radius of cross section for a pair of the colloidal and a liquid molecule. Then, $2\sigma'-\sigma$ is the diameter of the colloidal particle. The last two terms in the first line are given from the Langevin thermostat. The fluctuating force $\boldsymbol{\xi}_i$ is white noise with mean zero and the variance given by $\langle\xi_{ia}(t)\xi_{jb}(t')\rangle=2\delta_{ij}\delta_{ab}\delta(t-t')$ for $a, b=1,2,3$ denoting the components in 3 dimensions. The strength of the white noise, $\sqrt{\gamma \beta^{-1}}$, relates the dissipation coefficient $\gamma$ and the inverse temperature $\beta$ of the heat bath. This relation is called the Einstein relation or the fluctuation-dissipation relation, which guarantees that in the absence of the last term in Eq.~(\ref{reservoir}), the molecules reach an equilibrium with the Boltzmann distribution associated with unperturbed energy $E_{\textrm{hb}}^{(0)}$, which is given by $\sum_i m\mathbf{v}_i^2/2+\sum_{(i,j)} V_{\textrm{LJ}}(|\mathbf{x}_i-\mathbf{x}_j|;\epsilon,\sigma)$ where the subscript $(i,j)$ denotes the pair of molecules $i$ and $j$. Even in the presence of the interaction with the colloidal particle, the molecules approximately maintain equilibrium since the interaction force gives $\mathcal{O}(1)$-contribution to the dynamics while $E_{hb}^{(0)}\sim\mathcal{O}(N)$. This happens in experiments where the temperature of the liquid is kept well under the Brownian motion of a colloidal particle in the liquid. This can also be confirmed in MD simulations if the period of measurement is not too long. 

The equation of motion of the colloidal particle with the mass $M$ immersed in the liquid for position $\mathbf{x}$ and velocity $\mathbf{v}$ is given as 
\begin{equation}
M\dot{\mathbf{v}}=\mathbf{f}_{\textrm{app}}-\sum_{i}\boldsymbol{\nabla}V_{\textrm{LJ}}(|\mathbf{x}-\mathbf{x}_i|; \sigma')~
\label{colloidal}
\end{equation}
where $\mathbf{f}_{\textrm{app}}$ is an applied force acting exclusively on the colloidal particle, which is possible if an electromagnetic force is applied to a charged colloidal particle in an electrically neutral liquid. $-\sum_{i}\boldsymbol{\nabla}V_{\textrm{LJ}}(|\mathbf{x}-\mathbf{x}_i|; \sigma')$ is the sum of interaction forces exerted by molecules. It plays the equivalent role of dissipating and fluctuating forces in the Langevin dynamics. Heat is defined as work done by this force acting by molecules. 
 
We mimic an optical trap by harmonic force $-k\mathbf{x}$ with stiffness $k$, which is applied to confine the colloidal particle. Choosing $z$-axis to be perpendicular to the surface of the liquid, one can consider a non-conservative and linear force $\mathbf{f}_{\textrm{nc}}=-\left(
\begin{array}{cc}
0& \kappa_1\\
\kappa_2&0\\
\end{array}\right)
\left(\begin{array}{c} x\\y\end{array}\right)$ in horizontal direction to $x$-$y$ plane. 
Then the total applied force in horizontal direction is given by  
\begin{equation}
\mathbf{f}_{\textrm{app}}=-\mathsf{F}\cdot \mathbf{x}=-\left(
\begin{array}{cc}
k& \kappa_1\\
\kappa_2&k\\
\end{array}\right)
\left(\begin{array}{c} x\\y\end{array}\right)~.
\label{f_app}
\end{equation}
$\mathbf{f}_{\textrm{app}}$ becomes non-conservative for $\kappa_1\neq\kappa_2$, which is the source for nonequilibrium. It yields torque to produce nonzero circulation current even in the steady state. Divergence-less circulation current maintaining probability distribution is an important characteristics for nonequilibrium steady state~\cite{kwon-ao-thouless}. Energetically, nonzero current produces work and heat incessantly in time even in steady state~\cite{kwon,noh}. In reality there is a confining force in $z$-direction between the interfaces at top and bottom of the liquid. However, it usually depends on $z$ independent of $\mathbf{f}_{\textrm{app}}$, giving only simple equilibrium relaxation in $z$-space. For simplicity, we use the same harmonic force $-kz$. Non-equilibrium motion due to this $\mathbf{f}_{\textrm{app}}$ was studied via the overdamped Langevin equation and many interesting properties beyond the FT were found~\cite{kwon-ao-thouless,kwon,noh}, with which we will compare our MD simulation results. 

Experimentally, this kind of non-conservative force can be induced by time-dependent magnetic field $\mathbf{B}=B(t)\mathbf{k}$. The resultant force is $q\mathbf{v}\times \mathbf{B}-\partial\mathbf{A}/\partial t$ for charge $q$ and vector potential $\mathbf{A}$. For constant $\dot{B}$, the induced force is given by 
\begin{equation}
\mathbf{f}_{\textrm{ind}}=q\dot{B} t\left(
\begin{array}{cc}
0&1\\
-1&0\\
\end{array}\right)
\left(\begin{array}{c} v_x\\v_y\end{array}\right)+\frac{q\dot{B}}{2}\left(
\begin{array}{cc}
0& 1\\
-1&0\\
\end{array}\right)
\left(\begin{array}{c} x\\y\end{array}\right)~.
\end{equation}
In the overdamped limit with large friction coefficient $\gamma_c$ of the colloidal particle and short-time (or small $\dot{B}$) limit, the first term can be neglected compared to effective dissipating force $-\gamma_c\mathbf{v}$ in the regime $\gamma_c\gg q\dot{B}t$. In this regime, $\kappa_1=-\kappa_2=-q\dot{B}/2$.

In our work, we will carry out MD simulations for the non-equilibrium motion due to  $\mathbf{f}_{\textrm{nc}}$ in the part of Eq.~(\ref{f_app}) and confirm the results from the overdamped Langevin dynamics. The central quantities characterizing non-equilibrium motion are work $W$ done on the colloidal particle and heat $Q$ flowing into the heat bath, which are produced incessantly in time. The production rates of the two quantities are given by 
\begin{eqnarray}
\dot{W}&=&\mathbf{f}_{\textrm{nc}}\cdot\mathbf{v}=-\kappa_1v_x y-\kappa_2 v_y x ~,\label{work}\\
\dot{Q}&=&\sum_{i}\mathbf{v}\cdot\boldsymbol{\nabla}V_{\textrm{LJ}}(|\mathbf{x}-\mathbf{x}_i|; \sigma')~.
\label{heat}
\end{eqnarray}
The first law of thermodynamics is given by $dE/dt=\dot{W}-\dot{Q}$ where $E=m\mathbf{v}^2/2+k\mathbf{x}^2/2$.
We expect our study to serve as an alternative approach compared to the Langevin dynamics and to provide a basis to extend to more general cases for underdamped motion and long-time regime with $\gamma_c\sim q\dot{B}t$ or larger.

\section{Set-up for the simulation in overdamped limit}
To compare the results from MD simulations with those from the overdamped Langevin dynamics, we first estimate the friction coefficient $\gamma_c$ from simulations, which is a relative quantity between the colloidal particle and the liquid. In the following, we will present the simulation set-up to prepare a viscous liquid and a colloidal particle with large mass and size maintaining the overdamped limit, and how to estimate $\gamma_c$ between the two systems.

\subsection{Variables and parameters}

In our simulations, the liquid molecules and the colloidal particle are initially placed in a simulation box of size $L\times L \times L$. Periodic boundary conditions are imposed. Since the LJ potential is long-ranged, it will cost an enormous running-time to sum all the interactions. Instead, we set a cut-off distance $r_c$, above which the LJ potential can be truncated with negligible correction compared to thermal energies.  We adopt the parameters from the well-established MD studies on the LJ liquids~\cite{thompson,cieplak}. The particle number density is set by $0.81\sigma^{-3}$ and the temperature by $1.1\epsilon/k_B$, which was found to represent a compressed liquid above the melting temperature. $r_c=2.2\sigma$ was found for these parameters. We also use $r_c=2.2\sigma'$ for the interactions between the colloidal particle and a liquid molecule. We set the size of $L=21.46$ and about $9,000$ molecules are employed in simulations. 

We introduce dimensionless variables and parameters used for the simulation. We rescale $M/m\to M$, $\mathbf{x}/\sigma=\mathbf{x}$, $\beta^{-1}/\epsilon\to \beta^{-1}$, $t/\sqrt{m\sigma^2/\epsilon}\to t$. Then $m$, $\epsilon$, and $\sigma$ are set to unity, and  $\sigma'/\sigma$ to $\sigma'$. Given the above choice of temperature, $\beta^{-1}$ goes to $1.1$. In this rescaling, $V_{\textrm{LJ}}(|\mathbf{x}_i-\mathbf{x}_j|/\sigma;1)$ goes to $V_{\textrm{LJ}}(|\mathbf{x}_i-\mathbf{x}_j|; 1)$ and $V_{\textrm{LJ}}(|\mathbf{x}-\mathbf{x}_i|/\sigma; \sigma'/\sigma)$ to $V_{\textrm{LJ}}(|\mathbf{x}-\mathbf{x}_i|; \sigma')$. We also change $\gamma/\sqrt{m\epsilon/\sigma^2}\to \gamma$, $k/(\epsilon/\sigma^2)\to k$. $\kappa_1$ and $\kappa_2$ are set in the unit of $\epsilon/\sigma^2$ as $k$. The white noises having the unit of $t^{-1/2}$ are rescaled as $(m\sigma^2/\epsilon)^{-1/4}\boldsymbol{\xi_i}\to \boldsymbol{\xi_i}$. All changed variables and parameters become dimensionless. In the following sections, we will use the dimensionless variables and parameters, if not mentioned otherwise.

$\gamma$ is not a real coefficient of the friction exerted in the liquid, but a mathematical parameter of the Langevin thermostat. Then, we can choose $\gamma=1$ as long as the temperature of the heat bath is fixed from the Einstein relation, so the random force is written as $\beta^{-1/2}\boldsymbol{\xi}_i$ with  $\beta^{-1}=1.1$ in Eq.~(\ref{reservoir}). 

\subsection{Overdamped limit}

The experimental condition in most cases is consistent with the overdamped limit where the inertial effect is negligible. Our LJ-liquid itself is expected to have a large friction coefficient. However, a colloidal particle with lager size than that of liquid molecules will experience a much larger friction obeying the Stokes law: $\gamma_c=6\pi\eta R$ for the viscosity $\eta$ and the radius of the colloidal particle $R$. In the MD simulation, liquid is not continuous but discrete, so we expect $\gamma_c\propto\sigma'$ where $\sigma'$ is the radius of the cross section of a pair of the colloid and a liquid particle. There is restriction on the increase of $\sigma'$ in simulations. Under periodic boundary condition, the box size must be large enough to avoid possible unrealistic hydrodynamic interactions between the colloidal particle and its periodic replicas. A working criterion is $\sigma'/L < 1/10$, otherwise the chance of the colloidal particle interacting molecules near the boundary becomes too high. For $L=21.46$ used in our simulation, $\sigma'=1.5$ is about the maximal value we can obtain.  

From the point of view based on the Langevin dynamics, the equation of motion of the colloidal particle corresponding to Eq.~(\ref{colloidal}) is given as  $M\dot{\mathbf{v}}=\mathbf{f}_{\textrm{app}}-\gamma_c\mathbf{v}+\sqrt{\gamma_c\beta^{-1}}\boldsymbol{\xi}$, where $\gamma_c$ is the coefficient of the friction exerting on the colloidal particle in a viscous liquid and $\boldsymbol{\xi}$ is the same Gaussian noise as in Eq.~(\ref{reservoir}). Before investigating nonequilibrium motion, one can prepare the simulation set-up for the overdamped limit in an equilibrium situation. We consider an equilibrium case for $\mathbf{f}_{\textrm{app}}=-k\mathbf{x}$. In the overdamped limit, the fast-varying velocity is expected to relax more rapidly than the slowly-varying position. One can find the relaxation times of the velocity and position are given by $M/\gamma_c$ and $\gamma_c/k$, respectively. Then, the criterion for the overdamped limit is given by $M/\gamma _c\ll\gamma_c/k$, which can be achieved either by small $M$ or by large $\gamma_c$. In reality, $M$ is very large compared to the mass of liquid molecules, so $\gamma_c$ should be chosen large enough for the overdamped limit. 

\begin{figure}
\centering
\includegraphics*[width=\columnwidth]{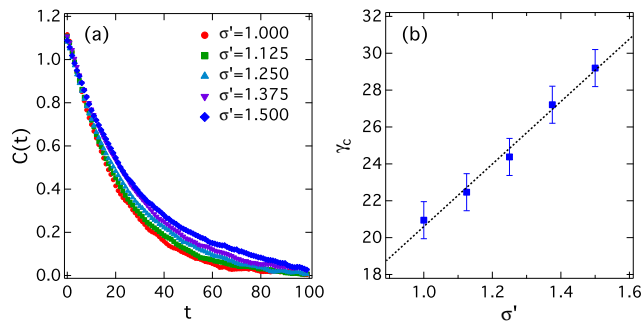}
\caption{(Color online) (a) The equilibrium correlation functions $C(t)$ in Eq.~(\ref{auto_corr_eq}) for $k=1$, $\kappa_1=\kappa_2=0$ for various values of $\sigma'$ show good agreements with $1.1 e^{-t/\gamma_c}$ for fitted values of $\gamma_c$. (b) The plot for $\gamma_c$ versus $\sigma'$ confirms the Stokes law well. }
\label{fig1} 
\end{figure}

We can estimate the friction coefficient $\gamma_c$ of the colloidal particle from the correlation function for the position in time which can be derived from the overdamped Langevin dynamics as 
\begin{equation}
C(t)=\langle x(t) x(0)\rangle=(\beta k)^{-1} e^{-kt/\gamma_c}~,
\label{auto_corr_eq}
\end{equation}
which becomes $1.1e^{-t/\gamma_c}$ for $\beta^{-1}=1.1$ and $k=1$. It is a special case with $\kappa_1=\kappa_2=0$ for the general formula derived in Eq.~(\ref{c11}). We repeat the simulations for $1\le \sigma'\le 1.5$. Figure.~\ref{fig1} confirms that $C(t)$ from the simulations fits the above theoretical equation very accurately and the estimated $\gamma_c$'s satisfy the Stokes law. $\gamma_c$ ranges from $20$ to $30$ for $1\le \sigma'\le 1.5$ and meets the criterion for the overdamped limit, $M/\gamma_c \ll\gamma_c/k$, for  $M=10$ used to mimic the large mass of the colloidal particle.  

\section{Simulation for Nonequilibrium}

We run simulations in discrete time steps with an interval $\Delta t=0.001$ and take $10^5$--$10^6$ samples produced by random noises in the Langevin thermostat and random initial conditions. The motion in $z$-direction is a simple equilibration process in an harmonic trap. The nonequilibrium motion occurs in $x$-$y$ plane where the applied force $-\mathsf{F}\cdot\mathbf{x}$ is given from Eq.~(\ref{f_app}) where
\begin{equation}
\mathsf{F}=\left(\begin{array}{cc} k &\kappa_1\\ \kappa_2 & k\end{array}\right)~. 
\label{F_matrix}
\end{equation}
The dynamics is shown to be stable so as to reach a new steady state if $\mathsf{F}$ is positive-definite ~\cite{kwon-ao-thouless,kwon}, i.e., $\textrm{Re}(k\pm\sqrt{\kappa_1\kappa_2})>0$. If $\mathsf{F}$ is symmetric with $\kappa_1=\kappa_2$, the force is conservative. Then, the steady-state distribution is  Boltzmann associated with the potential energy $1/2(kx^2 +ky^2+2\kappa_1 xy)$. If $\mathsf{F}$ is asymmetric with $\kappa_1\neq\kappa_2$, the force is nonconservative and the colloidal particle reaches a nonequilibrium steady state. 

We use $k=1$, $M=10$, and $\sigma'=1.5$. First, we carry out simulations with $\kappa_1=\kappa_2=0$ for a period much longer than the relaxation time $\gamma_c/k \sim 30$ in order for the colloidal particle to reach an equilibrium state. Then, we turn on a nonequilibrium protocol, $\kappa_1\neq\kappa_2$, which drives the colloidal particle out of equilibrium.  

\begin{figure}
\centering
\includegraphics*[width=\columnwidth]{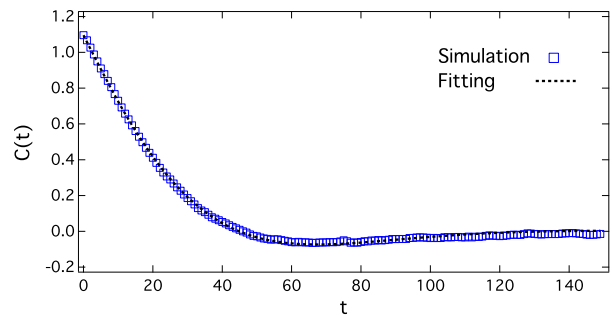}
\caption{(Color online) The correlation function $C(t)$ in nonequilibrium state for $\beta^{-1}=1.1$, $k=1$, $\kappa_1=-\kappa_2=1$, and $\sigma'=1.5$. The plot from the MD simulation agrees with that from Eq.~(\ref{auto_corr_neq}). Then, we estimate $\gamma_c\simeq 28.6$. }
\label{fig2}
\end{figure}

First, we examine the correlation function for the position in time and compare to the formula from the overdamped Langevin dynamics, given as
\begin{equation}
C(t)= \langle x(0)^2\rangle e^{-kt/\gamma_c}\cos\left[\frac{\sqrt{-\kappa_1\kappa_2}}{\gamma_c}t\right]~,
\label{auto_corr_neq}
\end{equation}
which is derived in detail in Eq.~(\ref{c11}). 
Since the colloidal particle is initially equilibrated for $\kappa_1=\kappa_2 =0$, $\langle x(0)^2\rangle=(\beta k)^{-1}$. Figure~\ref{fig2} shows a good agreement of the MD simulation with the above equation. We estimate $\gamma_c\simeq 28.6 $ for $\sigma'=1.5$, which agrees with the estimation from the simulation in equilibrium shown in Fig.~\ref{fig1}(b). 

Thermodynamic quantities characteristic to nonequilibrium are work and heat accumulated in time, which are expected to be produced persistently. The work and heat productions are found as
\begin{equation}
W=\sum_{n} \dot{W}_{n} \Delta t~,
~Q =\sum_{n} \dot{Q}_{n} \Delta t~,
\label{ave_production}
\end{equation}
where $\dot{W}_{n}$ ($\dot{Q}_{n}$) is the work (heat) rates at time step $n$, which are obtained from  Eqs.~(\ref{work}) and (\ref{heat}). The average values are found from $10^5$-$10^6$ samples.

\begin{figure}
\centering
\includegraphics*[width=\columnwidth]{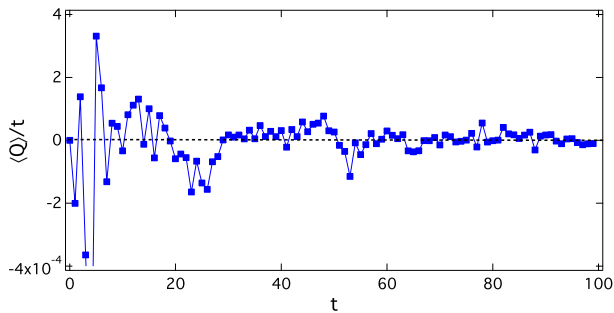}
\caption{(Color online) The simulation result for  of heat production in time for $k=1$ and $\sigma'=1$. $\langle Q/t\rangle$ goes to zero as time goes larger than the relaxation time $\gamma_c/k\sim 20$. $\gamma_c$ is given from the data in Fig.~\ref{fig2}. }
\label{fig3}
\end{figure}

Figure~\ref{fig3} shows the average rate of heat production in time for equilibration process where there is no work production for $\kappa_1=\kappa_2=0$. As expected, it decays as the system approaches equilibrium and there is no persistent production of heat. 

\begin{figure}
\centering
\includegraphics*[width=\columnwidth]{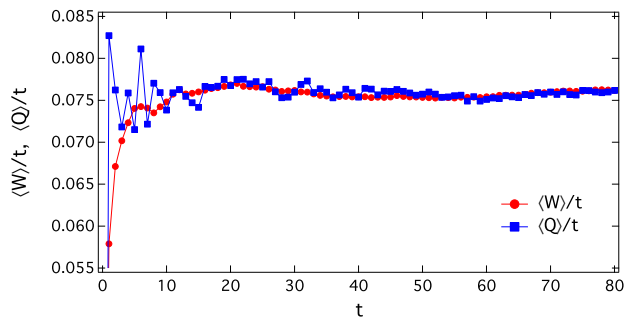}
\caption{(Color online) The average rates of work production and heat production in time, $\langle W\rangle/t$ and $\langle Q\rangle/t$, for $M=10$, $\sigma'=1.5$, $\beta^{-1}=1.1$, and $\kappa_1=-\kappa_2=1$. The two rates approach the same value for large $t$. The converging value  is approximately $0.076$ and the estimate value of $\gamma_c$ is equal to 29, which agrees very well with the values in Fig.~\ref{fig1}(b) and Fig.~\ref{fig2}.}.
\label{fig4}
\end{figure}

Figure~\ref{fig4} shows the average rates of work and heat production in time for $\kappa_1=-\kappa_2=1$. Time step $n$ is started after the simulation with $\kappa_1=\kappa_2=0$ is performed so that the colloidal particle is initially in equilibrium. The nonequilibrium steady state is characterized by the incessant productions of work and heat. Since $\langle E\rangle$ does not change in the steady state, the work and heat production become the same asymptotically for large $t$. In Fig.~\ref{fig4}, we observe that $\langle W\rangle\rangle/t$ and $\langle Q\rangle/t$ converge to the same value as the system approaches steady state. From a recent study via the overdamped Langevin equation~\cite{kwon}, It was found that $\langle W\rangle/t\to2[(\kappa_1-\kappa_2)/2]^2 \beta^{-1}/(k\gamma_c)$ as $t\to \infty$. In our dimensionless units, it is equal to $2.2/\gamma_c$ for $k=\kappa_1=-\kappa_2=1$, $\sigma'=1.5$, and $\beta^{-1}=1.1$. From the data in Fig.~\ref{fig4}, we estimate $\gamma_c\simeq 29$. All of the estimations of $\gamma_c$ from different methods presented in Figs.~\ref{fig1}, \ref{fig2}, and \ref{fig4} agree very well. 

\section{Work and heat distributions: Fluctuation Theorem}

The repeated measurements of work and heat produced during a period $t$ present large fluctuations over samples around the average values. The FT is a mathematical principle about the distribution functions for such fluctuations of thermodynamics quantities accumulated in time. It is convenient to consider the rates of work and heat, $w=W/t$ and $q=Q/t$, since $\langle W\rangle$ and $\langle Q\rangle$ increase linearly in large $t$, as shown in Fig.~\ref{fig4}. 

\begin{figure}
\centering
\includegraphics*[width=\columnwidth]{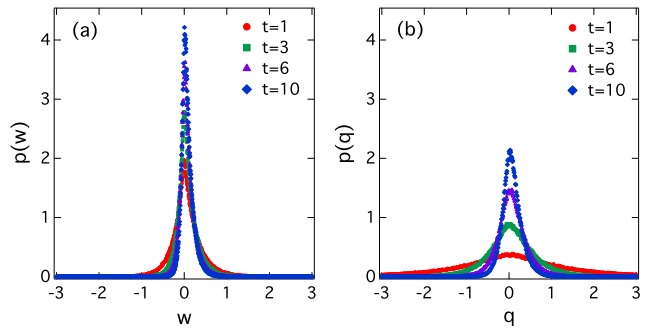}
\caption{(Color online) The distribution functions for (a) work and (b) heat produced for various times. The centers of the distributions approach the positive asymptotic value as in Fig.~\ref{fig3}. The two distributions show clear difference even for large $t$. The non-vanishing distribution for $w<0$ or $q<0$ for large $t$ shows that the thermodynamic second law can be violated in individual events.}
\label{fig5}
\end{figure}

We obtain the distribution functions for $w$ and $q$, given by $t^{-1}\sum_n\dot{W}_{n}\Delta t$, and $t^{-1}\sum_n\dot{Q}_{n}\Delta t$, respectively, from $10^5$--$10^6$ samples of the MD simulations. 
Figure~\ref{fig5} shows the distribution functions for $w$ and $q$. The two distributions show clear difference even for large $t$ where $\langle w\rangle$ becomes equal to $\langle q\rangle$. The thermodynamic second law reads: $\Delta S_{\textrm{tot}}=\Delta S_{\textrm{sys}}+Q/T\ge 0$ where $\Delta S_{\textrm{tot}}$ is the change in total entropy of the system and heat bath and $\Delta S_{\textrm{sys}}$  the change in system entropy, which leads to $Q/T\ge 0$ in steady state. In thermodynamic limit, since thermal fluctuations are negligible, the statistical average is not necessary. However, for small systems with large fluctuations, the second law can be shown to hold only in the average sense. From the FT for total entropy production: $\langle e^{-k_B^{-1}\Delta S_{\textrm{tot}}}\rangle=1$, one can get $\langle\Delta S_{\textrm{tot}}\rangle\ge 0$ using the Schwarz inequality. This means that the second law can be violated in individual events. In Fig.~\ref{fig5}, the probability of $w<0$ or $q<0$ does not vanish in long-time limit. 

\begin{figure}
\centering
\includegraphics*[width=\columnwidth]{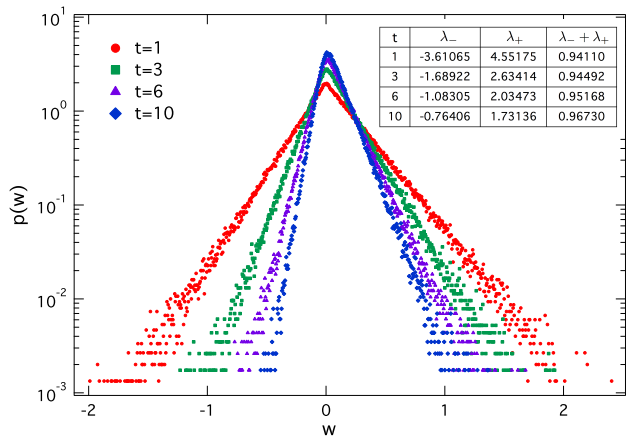}
\caption{(Color online) The semi-log plots of the distribution functions for work rate $w$ for various $t$ show exponential decays in both tails. 
In the inset, the table for the values of $\lambda_{\pm}$ from the slopes of the tails for various $t$ shows the Gallavotti--Cohen symmetry, $\lambda_++\lambda_-=1$.}
\label{fig6}
\end{figure}

In previous works~\cite{kwon,noh}, the distribution function $p(w)$ for $w$ was extensively investigated by means of the generating function $g(\lambda)=\langle e^{-\lambda\beta tw}\rangle$ where the bracket denotes the average over $w$. The so-called Gallavotti--Cohen (GC) symmetry was confirmed as $g(\lambda)=g(1-\lambda)$, which is known to be another representation of the FT. $g(\lambda)$ was found to be finite only for $\lambda_-<\lambda<\lambda_+$ and $\infty$ otherwise where $\lambda_+>0$ and $\lambda_-<0$. From the GC symmetry, $\lambda_+=1-\lambda_-$. It was found that the divergence of $g(\lambda)$ at $\lambda=\lambda_{\pm}$ determines an exponential decay in the tails of the work distribution. It was assumed that  
\begin{equation}
p(w)\sim \left\{ 
\begin{array}{ll}
|w|^{-\alpha} e^{t\lambda_- w}~,& w\to\infty \\
|w|^{-\alpha} e^{t\lambda_+ w}~, &  w\to-\infty
\end{array}\right.
\label{tail}
\end{equation}
Then, near $\lambda=\lambda_-<0$, the dominant contribution to $g(\lambda)$ comes from the integral in positive branch of $w$, given as 
\begin{eqnarray}
g(\lambda)&\sim&\int_{a>0}^\infty dW e^{-\beta t\lambda w}w^{-\alpha}e^{\beta t\lambda_-t}\nonumber\\
&\sim&(\lambda-\lambda_-)^{-1+\alpha}. 
\end{eqnarray}
The same divergence of $g(\lambda)$ near $\lambda=\lambda_+$ can be obtained from the divergent integral of $p(w)$ in negative $w$. The three types of tail behaviors were found as $\alpha=0~,1/2,~2$ depending on the form of $\mathsf{F}$. Our study belongs to the first type, while the other two occur for unequal diagonal elements of $\mathsf{F}$, corresponding to an anisotropic optical trap. We only restrict ourselves to isotropic optical trap and expect Eq.~(\ref{tail}). Figure~\ref{fig6} shows a clear view of the exponential decay of the distribution function for work, which is an indicator of the first type. The estimated values of $\lambda_{\pm}$ are shown to confirm the expected GC symmetry. 

\begin{figure}
\centering
\includegraphics*[width=\columnwidth]{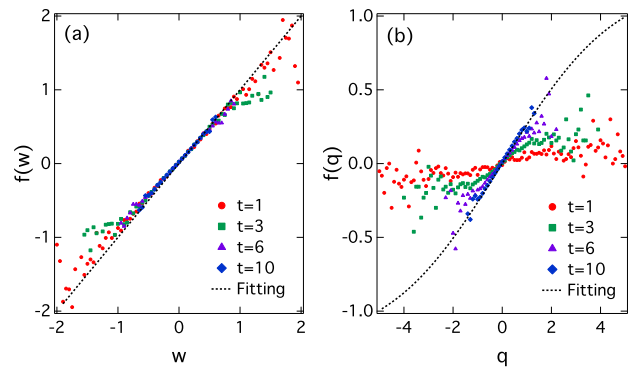}
\caption{(Color online) Define $f(u)=(\beta t)^{-1}\ln [p(u)/p(-u)]$ for $u=w,~q$. The examination of the FT for (a) the work production and (b) heat production from the MD simulations. In the panel (a), the data points are in good agreement with $f(w)=w$, which confirms the FT for work. In the panel (b), the guide lines, drawn using the fitting function $a~\!\textrm{erf}(bx)$, show that the FT for heat does not hold.}
\label{fig7}
\end{figure}

All the theoretical derivations of the FT are based on the Langevin equations or the master equations. We expect the FT to hold in the MD simulations as good as in the real experiments~\cite{wang,trepagnier,hummer,garnier,douarche,joubaud,liphardt,collin,hayashi,pak}. We examine the FT for work, which can be expressed in either integral form as $\langle e^{-\beta W}\rangle=1$ or detailed form as 
\begin{equation}
\frac{p(w)}{p(-w)}=e^{\beta tw}~.
\label{FT}
\end{equation}
In this work, we examine the detailed FT. In Fig.~\ref{fig7}, we plot $(\beta t)^{-1}\ln [p(w)/p(-w)]$, which is expected be $w$. The figure shows the FT for work holds for all times. 

We also examine the FT for heat. It was found that the FT for a certain time-accumulated quantity holds only if a proper initial distribution is given~\cite{seifert, esposito}. For the case of work, the initial distribution should be Boltzmann in equilibrium, as we prepare in MD simulations. For heat to present the FT, the initial distribution should be uniform ($\infty$-temperature distribution), which is hardly achievable in experiments.   
In steady state as $t$ goes to $\infty$, heat and work grow as $t$ while the energy difference remains finite. 
Moreover, the effect of the initial distribution decays exponentially in time. Therefore, one might expect that the FT for heat will approximately hold as $t$ goes large. However, equivalence between heat and work is only true in average sense. In Fig.~\ref{fig5}, the two distributions for large $t$ present clear difference in spite of the same average. The two path-dependent quantities are related by the first law as $Q=W-\Delta E$. By rare chances, however, $\Delta E$ can be comparable to work and heat for unbound state with $0\le E< \infty$, as in our case. If an initial (final) state is found from the tail with large energy far from the center of the distribution, $\Delta E$ can be negatively (positively) large so that it can significantly affect the positive (negative) tail of the heat distribution. Recent works revealed that the initial memory ever lasts in the positive tail of the heat distribution~\cite{farago,jslee}. This so-called boundary effect on the heat distribution due to $\Delta E$ was investigated in recent works~\cite{vanzon,puglisi,jd,jslee,kwangmoo}. 
Indeed, Figure~\ref{fig7}(b) shows the violation of the FT for heat even if $t$ increases. For large $t$, the slope of $f(q)=(\beta t)^{-1}\ln [p(q)/p(-q)]$ near the center of the distribution goes to unity while the tails remain nonlinear, as shown for the other nonequilibrium system~\cite{vanzon,kwangmoo}. 

\section{summary}

We investigate the motion of a colloidal particle moving in a liquid driven by a nonconservative force producing a torque by using MD simulations. The liquid composed of many molecules are designed to play the role of a heat bath at a fixed temperature by using the Langevin thermostat. The colloidal particle and the liquid molecules are designed to interact with each other via the LJ potentials. 

We mimic an experiment in the overdamped limit due to a large friction. We assign larger size to the colloidal particle than that of the liquid molecules. We estimate the values of the friction coefficient $\gamma_{\textrm{c}}$ from three methods: the equilibrium and nonequilibrium correlation functions for position in time and the asymptotic production rate of work for infinite $t$. The three values are found to be identical with high accuracy and to hold the Stokes law, which assures us of the reliability of our MD simulations.  

After running the simulations for a sufficiently long time in the absence of nonconservative force for the colloidal particle to reach equilibrium, we turn on a nonconservative force with $\kappa_1\neq\kappa_2$ and perform measurements for work and heat produced for period $t$. From $10^5$--$10^6$ samples due to random numbers used for the Langevin thermostat and random initial conditions, we obtain the distribution functions for work and heat.  We observe that the second law of thermodynamics can be violated in individual events while satisfied in average. The distribution functions for work for various $t$ are found to decay exponentially in the both tails of the distribution. We confirm the Gallavotti--Cohen symmetry inherent in the tails. We also confirm the detailed FT to hold for work, but not to hold for heat, as expected for nonequilibrium processes starting initially from equilibrium. We discuss the boundary effect on the heat distribution to yield the violation of the FT for heat.

We are interested in the MD simulations for long-time regime with $\gamma_c\sim q\dot{B} t$ or larger that yields a nonvanishing time-dependent Lorentz force in addition to the torque-driving force in this work. It will be 
interesting to investigate the nonequilibrium motion beyond the overdamped limit by increasing the mass of the colloidal particle or decreasing the friction coefficient and other examples driven by different nonequilibrium sources.

\begin{acknowledgments}    
This work was supported by the Basic Science Research Program through
the NRF Grant No.~2013R1A1A2011079 (CK) and 2015R1D1A1A09057469 (YJ), and also by the Korea Institute of Science and Technology Information with supercomputing resources including technical support (KSC-2015-C1-001). 
\end{acknowledgments}

\appendix
\section{\label{appendix}The derivation of Correlation Matrix}
The equation of motion of the colloidal particle in the overdamped limit is given by the Langevin equation:
\begin{equation}
\dot{\mathbf{x}}(t)=-\gamma_c^{-1} \mathsf{F}\cdot\mathbf{x}(t)+\boldsymbol{\xi}(t),
\label{overdamped_langevin}
\end{equation}
where $\mathsf{F}=\left(\begin{array}{cc} k&\kappa_1\\ \kappa_2 & k\end{array}\right)$ and $\langle\xi_a(t)\xi_b(t')\rangle=2\beta^{-1}\gamma_c^{-1}\delta_{ab}\delta(t-t')$ for $a,b=1,2$. Let $\mathsf{C}(t)$ be the correlation matrix with components $C_{ab}(t)=\langle x_a(t) x_b(0)\rangle$. Then, one can find $d\mathsf{C}(t)/dt =-\gamma_c^{-1}\mathsf{F}\mathsf{C}(t)$ by multiplying Eq.~(\ref{overdamped_langevin}) by $\mathbf{x}(0)$, where $\langle \xi_a(t) x_b(0)\rangle=0$ is used. The solution is given by $\mathsf{C}(t)=e^{-\gamma_c^{-1}\mathsf{F} t}\mathsf{C}(0)$. 

$\mathsf{F}$ has two eigenvalues $\lambda_{\pm}=k\pm \sqrt{\kappa_1\kappa_2}$ and orthonormalized left (right) eigenvectors $\langle\pm|$ ($|\pm\rangle$) given as 
\begin{equation}
\langle\pm|=\frac{1}{\sqrt{2}}(1,\pm\sqrt{\kappa_1/\kappa_2})~,~~|\pm\rangle=\frac{1}{\sqrt{2}}\left(\begin{array}{c}1\\ \pm\sqrt{\kappa_2/\kappa_1}\end{array}\right).
\end{equation}
Then, one can show $\mathsf{F}=\lambda_{+}|+\rangle\langle +|+\lambda_{-}|-\rangle\langle-|$ with the property $\mathsf{F}^n=\lambda_{+}^n|+\rangle\langle +|+\lambda_{-}^n|-\rangle\langle-|$. Using this property, one can obtain $e^{-\gamma_c^{-1}\mathsf{F}t}=e^{-\gamma_c^{-1}\lambda_{+}t}|+\rangle\langle +|+e^{-\gamma_c^{-1}\lambda_{-}t}|-\rangle\langle -|$, which gives 
\begin{equation}
e^{-\gamma_c^{-1}\mathsf{F}t}=\frac{1}{2}\left(
\begin{array}{cc}  a_+ & \sqrt{\frac{\kappa_1}{\kappa_2}}a_-\\  \sqrt{\frac{\kappa_2}{\kappa_1}}a_-&a_+\end{array}\right),
\end{equation}
where $a_+=e^{-\gamma_c^{-1}\lambda_{+}t}+e^{-\gamma_c^{-1}\lambda_{-}t}$ and $a_-=e^{-\gamma_c^{-1}\lambda_{+}t}-e^{-\gamma_c^{-1}\lambda_{-}t}$.
In our work, we consider nonequilibrium protocols $\kappa_1\neq \kappa_2$ which are turned on at $t=0$ as the system is initially in equilibrium for $\kappa_1=\kappa_2=0$ with $C_{11}(0)=C_{22}(0)=(\beta k)^{-1}$ and $C_{12}(0)=C_{21}(0)=0$. In particular, we consider the case for $\kappa_1\kappa_2<0$ and find
\begin{eqnarray}
C_{11}(t)&=&C_{11}(0)e^{-\gamma_c^{-1}kt}\!\!\cos\left[\frac{\sqrt{-\kappa_1\kappa_2}}{\gamma_c}t\right],\label{c11}\\
C_{12}(t)&=&C_{11}(0)\sqrt{\left|\frac{\kappa_1}{\kappa_2}\right|}e^{-\gamma_c^{-1}kt}\sin\left[\frac{\sqrt{-\kappa_1\kappa_2}}{\gamma_c}t\right],\\
C_{21}(t)&=&C_{11}(0)\sqrt{\left|\frac{\kappa_2}{\kappa_1}\right|}e^{-\gamma_c^{-1}kt}\sin\left[\frac{\sqrt{-\kappa_1\kappa_2}}{\gamma_c}t\right],
\end{eqnarray}
with $C_{22}(t)=C_{11}(t)$.
We examine $C_{11}(t)$ by MD simulation for an equilibrium case with $\kappa_1=\kappa_2=0$ and a nonequilibrium case with $\kappa_1\neq\kappa_2=0$, given in Eq.~(\ref{auto_corr_eq}) and Eq.~(\ref{auto_corr_neq}), respectively.

\end{document}